\documentclass[12pt]{article} 
\usepackage{amsmath} 
\usepackage{amsthm} 
\usepackage{amssymb}    
\usepackage{graphicx} 
\usepackage{multicol} 
\usepackage[dvips,letterpaper,margin=0.75in,bottom=0.5in]{geometry}  
\usepackage{fullpage}
\usepackage{graphicx}
\usepackage{epsfig}
\usepackage{subfigure}
\usepackage{latexsym}
\usepackage{times}
\usepackage{amsmath}

\makeatletter 

\renewcommand{\maketitle} 
{ \begingroup \vskip 10pt \begin{center} \large {\bf \@title}
        \vskip 10pt \large \@author \hskip 20pt \@date \end{center}
  \vskip 10pt \endgroup \setcounter{footnote}{0} }

\begin{document}

\centerline{\Large \bf \underline {Active Matter}}
\vskip 1cm
\centerline{\large Gautam I. Menon}
\centerline{\large The Institute of Mathematical Sciences}
\centerline{\large  CIT Campus, Taramani, Chennai 600 113 INDIA}
\begin{abstract}
The term \emph {active matter} describes
diverse systems, spanning macroscopic (e.g.
shoals of fish and flocks of birds) to microscopic
scales (e.g. migrating cells, motile bacteria
and gels formed through the interaction of nanoscale
molecular motors with cytoskeletal filaments within
cells). Such systems are often idealizable in terms
of collections of individual units, referred to as
active particles or self-propelled particles, which
take energy from an internal replenishable energy
depot or ambient medium and transduce it into useful work performed
on the environment, in addition to dissipating a fraction
of this energy into heat. These individual units may interact both directly
as well as through disturbances propagated via the medium in which they are immersed.
Active particles can exhibit remarkable collective
behaviour as a consequence of these interactions,
including non-equilibrium phase transitions between
novel dynamical phases, large fluctuations violating
expectations from the central limit theorem and
substantial robustness against the disordering effects
of thermal fluctuations.  In this chapter, following
a brief summary of experimental systems which may be
classified as examples of active matter,
I describe some of the principles which underlie the modeling of
such systems.
\end{abstract}

\section{Introduction to Active Fluids}
\label{sec:intro}

Anyone who has admired the intricate dynamics of
a group of birds in flight or the coordinated,
almost balletic maneuvers of a school of swimming
fish can appreciate the motivation for the study of
``active matter'':  How do individual self-driven units, such
as wildebeest, starlings, fish or bacteria, flock
together, generating large-scale, spatiotemporally
complex dynamical patterns \cite{vicsekPRL95}? What are the rules which
govern this dynamics and how do the principles of
physics constrain the behaviour of each such unit? Finally,
what are the simplest possible models for such
behaviour and is there any commonality to the
description of these varied problems~\cite{tonertuPRL95,tonertuPRE98,sriramANNPHYS05}?

Describing these diverse problems in terms of
individual ``agents'' which evolve via a basic
set of update rules while interacting with other agents
provides a general way of approaching a large number
of unrelated problems. These include the description of
the propagation of infectious diseases in a population,
the seasonal migration of animal populations, the
collective motion and coordinated activities of groups
of ants and bees and the swimming of shoals of fish~\cite{sriramANNPHYS05}.
Of these problems, the subset of problems involving
agents whose \emph{mechanical} behaviour at a scale larger
than the individual agent must be
constrained by local conservation laws, such as the
conservation of momentum, forms a special
class~\cite{aditiPRL02}. It is these systems that are the primary focus
of this chapter.

Generalizing from the examples above, a tentative
definition of active matter might be the following:
\emph {Active matter is a term which describes a material
(either in the continuum or naturally decomposable into
discrete units), which is driven out of equilibrium
through the transduction of energy derived from an
internal energy depot or ambient medium into work performed on the
environment.}
Such systems are generically capable of emergent
behaviour at  large scales.  The hydrodynamic
description of such problems should be equally
applicable to situations where the granularity of the
constituent units is not resolved, but which are
rendered non-equilibrium in a qualitatively similar way~\cite{sriramANNPHYS05,sriramNJP07}.

What differentiates active systems from other classes
of driven systems (sheared fluids, sedimenting
colloids, driven vortex lattices etc.) is that the
energy input is internal to the medium (i.e. located on
each unit) and does not act at the boundaries or via
external fields. Further, the direction in which each unit moves
is dictated by the state of the particle and not by the 
direction imposed by an external field.
\begin{figure}[htp]
\centering
\includegraphics[scale=.65]{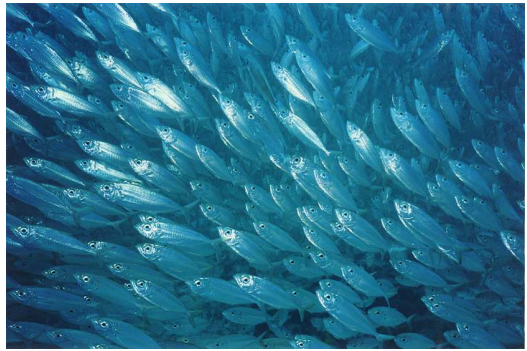}
\caption{A school of fish, illustrating the tendency towards parallel 
alignment while swimming. Picture courtesy Prof. R. Kent Wenger}
\label{school}       
\end{figure}

\begin{figure}[htp]
\centering
\includegraphics[scale=.55]{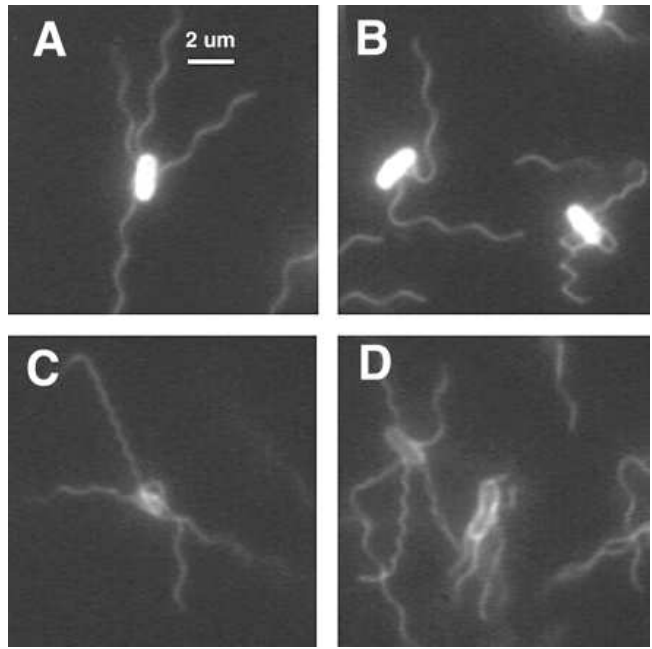}
\caption{Cells of \emph{Escherichia coli} labeled with a dye and examined in 
a fluorescence microscope. From these images, the 
behaviour of the flagellae of individual cells can be studied 
as \emph{E. coli} swims~\cite{turner2000}. Picture courtesy Prof. Howard Berg}
\label{ecoli}       
\end{figure}

Such  individual units are, in general, anisotropic, as in Fig.~\ref{school}.
Collections of such units are thus capable
of exhibiting orientationally ordered states.
A canonical example of an ordered state of orientable
units obtained in thermal equilibrium is the nematic
liquid crystal, in which anisotropic particles align
along a common axis. Active systems of such units
can, in addition, exhibit directed motion along this
axis; a concrete example is illustrated in Fig.~\ref{school}. In the context of active particles in a fluid,
the terminology ``swimmers'' or ``self-propelled
particles'' is often used, while the terms ``active
nematic'' or ``living liquid crystals'' occur in the
discussion of the orientationally ordered collective
states of active particles~\cite{sriramNJP07}.

Why study active systems?  For one, such systems
can display phases and phase transitions absent in
systems in thermal equilibrium~\cite{tonertuPRL95,tonertuPRE98}.  For another, active matter often exhibits unusual mechanical properties,
including strong instabilities of ordered states
to small fluctuations. Such instabilities may be
\emph{generic} in the sense that they should appear
in any hydrodynamic theory which enforces momentum
conservation and includes the lowest order contribution
to the system stress tensor arising from activity~\cite{aditiPRL02}.  
Fluctuations in active systems are generically
large, often deviating qualitatively  from the predictions
of simple arguments based on the central limit theorem~\cite{PHYSICA02aditi,tonertuPRL95,tonertuPRE98,EPL03sriram}. 
In common with a large number of related
non-equilibrium systems, some examples of active
matter appear to be self-tuned to the vicinity of a
phase transition, where response can be anomalously
large and fluctuations dominate the average behaviour.

The study of active matter spans many scales. The
smallest scales involve the modeling of individual
motile organisms, such as fish or individual
bacteria or even nanometer-scale motor proteins such
as kinesin or dyneins
which move in a directed manner along cytoskeletal
filaments. Hydrodynamic descriptions of large numbers
of motile organisms operate at a larger length scale,
averaged over a number of such swimmers that is
large enough for the granularity at the level of the
individual swimmer to be neglected, while still allowing
for the possibility of spatial fluctuations  at an intermediate
scale.

Finally, all \emph{living} matter is matter out
of thermodynamic equilibrium. The source of this
non-equilibrium is, typically, the hydrolysis of a NTP
(nucleoside tri-phosphate, such as ATP or GTP) molecule
into its di-phosphate form, releasing energy. This
energy release can drive conformational changes in a
protein, leading to mechanical work. Biological matter
is thus generically internally driven, precisely
as demanded by our definition of active matter.
Understanding the general principles which govern
active matter systems could thus provide ideas,
terminology and a consistent set of methods for the
modeling of the mechanical behaviour of living, as
opposed to dead, biological matter. This is perhaps
the most substantial motivation for the study of
active matter.

The outline of this chapter is the following: Section~\ref{sec:examples}
provides some examples of matter which can be
classified as active, using our definition.
Section~\ref{sec:individual} summarizes
results on a simple model for an individual
swimmer and summarizes necessary ingredients for a 
coarse-grained description of a large number of
swimmers. Section~\ref{sec:activelc} begins by
illustrating the derivation of the equations of motion
and the stress tensor of
a simple fluid and then goes on to illustrate how this
derivation may be extended to describe fluids with orientational
order close
to thermal equilibrium. It then goes on to discuss 
how novel contributions
to the fluid stress tensor can arise from active
fluctuations, deriving an equation of motion for small
fluctuations above a pre-assumed nematic state. These
calculations lead to  predictions for the
rheological behaviour of active systems and the demonstration of
the instability of the ordered nematic (or polar) state to small
fluctuations. In Sect.~\ref{sec:activegel},
the theory of active gels is summarized briefly and some
results from this formalism are outlined. Section~\ref{sec:conclusions} 
provides a brief summary.

\section{Active Matter Systems: Some Examples}
\label{sec:examples}

The subsections below list a (non-exhaustive)
set of examples of active matter systems. The
large-scale behaviour of some of these systems is the
focus of later sections.
\subsection{Dynamical Behaviour in Bacterial Suspensions} 
Dombrowski et.al. study the
velocity fields induced by bacterial motion at the
bottom of sessile and pendant drops containing the
bacterium \emph{B. subtilis}~\cite{dombrowski}. These
flows reflect the  interplay between bacterial
chemotaxis and buoyancy effects which together act to
carry bioconvective
plumes down a slanted meniscus, concentrating
cells at the drop edge (sessile) or at the drop bottom
(pendant).  The motion exhibited by groups of bacteria
as a result of this self-concentration  is large-scale
at the level of individual bacteria, exhibiting
vortical structures and other complex  patterns as a
consequence of the hydrodynamic interaction between
swimming bacteria. Thus, these experiments highlight
the crucial role of hydrodynamics, coupled with the
motion of individual active swimmers, in generating
self-organized, large-scale dynamical fluctuations in
the surrounding fluid.

\subsection{Mixtures of Cytoskeletal Filaments and
Molecular Motors} 

The influential experiments of Nedelec and collaborators combine a limited number of cytoskeletal and motor elements -- microtubules, kinesin
complexes and ATP -- in a two-dimensional geometry
in vitro finding remarkable self-organized patterns evolving
from initially disordered configurations~\cite{nedelec}.
These include individual  asters and  vortices as
well as disordered arrangements of such structures,
in addition to bundles and disordered states at varying values of
motor densities.  These are self-organized structures, 
large on the scale of the individual cytoskeletal filament and motor, which require ATP (and thus are non-equilibrium) in order to form and
be sustained. Several hydrodynamical approaches to this problem have been proposed, as in Ref.~\cite{leePRE01,sumithraPRE04,aranson}. 
These experiments and associated theory 
are summarized in Ref.~\cite{NCB}.

\subsection{Layers of Vibrated Granular Rods} 

The experiments of Narayanan et.al. take elongated
rod-like copper particles in a quasi-two-dimensional
geometry, and vibrate them vertically in a
shaker~\cite{vijayN}. This agitated monolayer of
particles is maintained out of equilibrium by the
shaking, which effectively acts to convert vertical motion
into horizontal motion via the tilting of the rod. The particles are back--front
symmetric, and are thus nematic in character.
These experiments see large, dynamical regions
which appear to fluctuate coherently. Similar behaviour is
predicted in theories
of flocking behaviour induced purely by increasing
the concentration in dense aggregates of particles
held out of equilibrium~\cite{vicsekPRL95,sriramANNPHYS05}. These experiments provide
a particularly striking signature of non-equilibrium
steady state behaviour in ``flocking'' systems: the
presence of number fluctuations in such systems which
scale anomalously with the size of the region being
averaged, i.e., with a power of $N$ which exceeds
the central limit prediction of a $\sqrt{N}$ behaviour
of number fluctuations~\cite{sriramANNPHYS05}.  
\subsection{Fish Schools }
Experiments of Makris and collaborators use remote
sensing methods on a continental scale to access the structure and dynamics of
large-scale shoals of fish, containing an order of a few
million individuals~\cite{makrisSCIENCE06}. Apart from
the relatively rapid time-scale for the reorganization
of the shoal -- around 1--10 min in their experiments
-- these experiments also see evidence for ``fish
waves''; propagating internal disturbances within the
shoal which occur at relatively regular intervals,
representing disturbances at scales far larger
than that of the individual fish. The speeds of
such waves are larger, by around a factor of 10,
than the velocities of the swimming fish. They appear
to represent ``locally interconnected compaction
events'' similar to the Mexican waves exhibited through the
coordinated motion of spectators in stadia.  It is interesting that
similar wave-like excitations are predicted in the
hydrodynamic theories of Refs.~\cite{tonertuPRE98,aditiPRL02},
where they involve waves of concentration and splay~\cite{tonertuPRE98}
or splay--concentration and bend~\cite{aditiPRL02}.

\subsection{Bird Flocks} 

The STARFLAG collaboration has imaged large flocks of starlings
(between $10^3$ and $10^4$ individuals at a time), using
computer-aided imaging techniques to understand the
dynamics of individual birds and how this dynamics is
influenced by the spatial distribution and
behaviour of neighbouring birds. Surprisingly, and
contrary to what might have been naively expected,
birds appear to adjust to the motion of the flock
by measuring the behaviour of topologically (and
not metrically) related neighbours~\cite{starflag}.
Thus, at each instant, each bird appears to be comparing
its instantaneous position and velocity to those of the
5--7 birds closest to it, making the adjustments required to
maintain the coherence of the flock.
This strategy appears to have the advantage that
reducing the density of the flock
should not then impact the coherence of the flock,
since only topological and not metrical relationships
are involved, a fact that calls into question the gradient
expansions favoured by most theoretical work which
represents flocking behaviour by coarse-grained
equations of motion for a few hydrodynamic fields.

\subsection{Marching Behaviour of Ants
and Locusts} Ian Couzin's group at Princeton has
investigated the transition to marching behaviour
in locusts. Swarms of the desert locust, \emph{
Schistocerca gregaria}, in their non-flying form,
can exist in a relatively solitary individualistic
state as well as a gregarious collective state. In
the ``gregarious'' state, such locusts can form huge
collective marching units which forage all vegetation
in their path. This has a  huge social and economic
impact on humans, affecting the livelihood of one
in ten people on the planet in plague years. These
experiments provide an experimental example of the
collective transition in simple computational models
of flocking behaviour. From more  recent work from
this group, it appears, somewhat unusually, that the
transition is induced by ``cannibalistic'' behaviour,
in which a locust  is successful in
biting the rear-quarters of the  locust immediately in
front~\cite{bazazi08,buhlSCIENCE06}.

\subsection{Listeria monocytogenes Motility}
The bacterium \emph{Listeria monocytogenes} is
a simple model system for cell motility, which derives
its ability to move from the polymerization of actin,
leading to the formation of a ``comet tail''
emerging from the rear of the bacterium~\cite{theriot}. Motility appears to
arise from the deformation of the gel formed by the actin and cross-linking proteins as a consequence of continuing polymerization,
which results in a propulsive force on the bacterium. The
non-equilibrium comes from ATP-driven actin
polymerization.  Interestingly, many features of the experiment
can be reproduced in in vitro systems where actin polymerization
is initiated at the surface of specially treated beads, which
then exhibit symmetry-breaking motility~\cite{actinbead}.

\subsection{Cell Crawling}
There is a vast and intriguing literature on cell crawling on substrates~\cite{bershadsky,bray,grulerEPJB99,verkhovsky}.
Such crawling appears to have four basic steps: the extension of
cellular protrusions, the attachment to the substrate at the leading
edge, the translocation of the cell body and the detachment at
the rear. Cell crawling appears to be largely mediated by a meshwork of
actin in gel form, whose fluidity is actively maintained through
the action of myosin motors and other associated 
proteins~\cite{bershadsky,verkhovsky}. Interestingly, motility has
also been observed in nucleus-lacking cell fragments excised by lasers,
over a period of several hours.

\subsection{Active Membranes}
Experiments on fluctuating giant vesicles containing bacteriorhodopsin (BR) 
pumps reconstituted in a lipid bilayer indicate that the light-driven proton pumping
activity of BR amplifies membrane shape fluctuations\cite{mannevillePRE01,mannevillePRL99}. The BR pumps
transfer protons in one direction across the membrane as they
change conformation upon  excitation by light of a specific wavelength. 
These experiments have been described in terms of a non-equilibrium ``active'' 
temperature\cite{mannevillePRE01,mannevillePRL99}. Hydrodynamic theories which interpret the experiments {\it via}
a description of a membrane with a density of embedded diffusing dipolar force 
centres,  {\it i.e.} as an ``active membrane'',  have also been studied in some detail.  It is interesting that several
developments in the theoretical description of generic active matter reflect  ideas first introduced in the context of active
membranes\cite{prostEPL96,sriramPRL00,sumithraPRE02}. A detailed study of a biologically relevant model
of an active membrane system which contains a large number of references to the 
literature is available in Ref.~\cite{lacoste}.


\section{The Swimmer: Individual and Collective}
\label{sec:individual}

The examples provided in the previous section are
indicative of some of the diversity exhibited by
systems which fall under the general category of active
matter. These can broadly be classified into systems
in which the role of mechanical conservation laws
(e.g. for momentum) are important and systems
in which, typically, only the conservation of
number is relevant. 

The  flocking model of Vicsek and collaborators~\cite{vicsekPRL95},
recast in terms of  hydrodynamic equations of motion
by Toner and Tu~\cite{tonertuPRL95,tonertuPRE98,sriramANNPHYS05}, is an
agent-based model in which cooperativity is driven
by a density dependent interaction which tends to
orient individual units in the direction in which
their neighbours move. The environment -- the ground
and vegetation for the moving locusts, for example,
provides merely the background in this case and 
has no other dynamical significance. In the case
of the flock of starlings the communication
between individual starlings does not appear to be
primarily via the medium, but through visual contact. 

The case of the motile bacterium and the swimming fish,
on the other hand, are cases where the medium plays
an important role in transferring momentum to and
from the swimmers and in determining the interactions 
between swimmers. This case will be discussed below,
first for the situation of the individual swimmer
and then for collections of swimmers, modeled via a
hydrodynamic approach.

\subsection{The Individual Swimmer}
The central idea in the modeling of individual swimmers
is that the system must be force-free, when averaged over
lengthscales larger than the characteristic dimension
of the swimmer~\cite{sriramNJP07, aparnaPNAS09}. This is
a consequence of Newton's third law, which imposes 
that the force exerted by the swimmer
on the fluid must be equal and opposite to the force
exerted by the fluid on the swimmer.  Thus, if one
considers the swimmer as a source of forces locally
within the fluid which act upon the fluid, such a
source cannot have a monopole component but may
have a dipole (or higher multipole) component.

Depending on the character of the swimmer, more
distinctions are possible~\cite{aparnaPNAS09,sriramNJP07}. Contractile swimmers or pullers (such as bacteria propelled by flagella at the
head of the organism) pull fluid in along the long
axis and push it out along an axis normal to their
midpoint. Tensile swimmers  or pushers push fluid
out along their long axis and pull fluid in along
the midpoints. They are propelled from the rear, 
justifying the terminology of pushers. 

Consider a simple model for a microscopic swimmer. In
the Stokes limit, time does not enter these equations
explicitly and the velocity field is completely
specified by the boundary conditions imposed on the
flow. (The  use of the Stokes limit is generically justified
in the case of bacteria, where characteristic Reynolds numbers
at the scale of a single particle
are of the order of $10^{-4}$ or smaller.) Reversing the velocity 
field at the boundaries should retrace the velocity field configuration,
implying that the trajectory assumed by the swimmer
in its configuration space cannot be time-reversal
invariant. The helical or ``corkscrew-like'' motion of
the flagellae of the bacterium \emph{E. coli} (see Fig.~\ref{ecoli}), discussed
by Purcell~\cite{purcell}, provides a particularly attractive example
of how the limitations on directed motion at the low
Reynolds numbers required for the Stokes limit approximated
to be valid can be overcome.

Averaging over multiple strokes of the swimmer
simplifies the description: the broken temporal
symmetry required for translation in a flow governed
by the Stokes equations can be replaced by a broken
spatial symmetry. This singles out the direction of
motion of the swimmer.  Following a model introduced
recently by Baskaran and Marchetti (whose notations
and treatment we follow closely in this section),
the swimmer is modeled as an asymmetric, rigid
dumbbell~\cite{aparnaPNAS09}. This dumbbell consists of two differently
sized spheres, of
radii $a_L$ (large) and $a_S$ (small), forming the head and the tail
of the swimmer.  The length $\ell$ of the swimmer is
the length between the two centers.  The orientation
of the swimmer is given by the unit vector $\hat{v}$,
drawn from the smaller sphere to the larger sphere.

Thus, the equations of motion of the two spheres, 
with locations  ${\bf r}_{L\alpha}$ and ${\bf r}_{R\alpha}$, are
given by
\begin{eqnarray}
\partial_t {\bf{r}}_{L\alpha} &=& {\bf u}({\bf r}_{L\alpha}) \nonumber \\
\partial_t {\bf r}_{S\alpha} &=& {\bf u}({\bf r}_{S\alpha}) \;,
\end{eqnarray}
where the constant length between the two spheres is imposed by
the constraint that ${\bf r}_{L\alpha} - {\bf r}_{S\alpha} = \ell \hat v_\alpha$
and the velocity field is given by ${\bf u}({\bf r})$.
The no-slip condition at the surfaces of the spheres requires that the
sphere move with the velocity of the fluid next to it.

The velocity field obeys the Stokes equation and is constrained by incompressibility, 
\begin{eqnarray}
\eta \nabla^2 \bf u(r) &=& \nabla p + {\bf F}_{active} + {\bf f}_{Noise}, \nonumber \\
\nabla \cdot \bf u(r) &=& 0,
\label{stokes}
\end{eqnarray}
where the ``force'' terms which enter on the right hand side are
composed of a term which is associated purely with activity (i.e.
vanishes in thermal equilibrium) as well as of a second ``noise''
term modeling the fluctuations in fluid velocity arising from
purely thermal fluctuations. Such noise terms do not conventionally
enter the Navier-Stokes equations, but must generically be included 
in a coarse-grained description. Detailed expressions for these
forces are available in Ref.~\cite{aparnaPNAS09}.

To solve the Stokes equations,
we insert a delta function force $\delta(\bf r)\bf F$ on the right hand
side, representing a fundamental source term from which more complex
force configurations can be constructed. 
In Fourier space, the incompressibility condition is imposed as
${\bf q} \cdot \bf u(q) = 0$,
while the Stokes equations are $
-\eta q^2 {\bf u(q)}  + i {\bf q} p(\bf{q}) = \bf F(q)$.
This then gives
\begin{equation}
{\bf u(r)} = \frac{1}{\eta q^2} \left({\bf F} - \frac{ ({\bf q} \cdot {\bf F}) {\bf q}}{q^2}\right)\;,
\end{equation}
with 
\begin{equation}
{p(\bf q)} = i \frac{{\bf q} \cdot {\bf F}}{q^2}.
\end{equation}
These are inverted by 
\begin{equation}
{\bf u(r)} = \frac{1}{8 \pi\eta }\left ( \frac{\bf \delta}{r}
+ \frac{\bf r \bf r}{r^3} \right ) \cdot {\bf F}.
\end{equation}
The ${\boldsymbol \delta}$ symbol is the unit tensor. In a Cartesian basis of ${\bf i,j,k}$ it is
\begin{equation}
{\boldsymbol \delta} = {\bf i}{\bf i} + {\bf j}{\bf j} + {\bf k}{\bf k} 
\end{equation}
The quantity acting on the force on the right hand side 
is the Oseen tensor, defined through
\begin{equation}
{\cal O}_{ij}({\bf r}) = \left (\delta_{ij} + {\hat r}_i{\hat r}_j \right )/8\pi\eta r
\end{equation}
for $|r| > a_{L,S}$, with ${\hat r} = {\bf r}/|{\bf r}|$ a unit vector.
The Stokes equations are solved by the superposition
\begin{equation}
{u_i(\bf r)} = f\sum_\alpha \left [{\cal O}_{ij}({\bf r} - {\bf r}_{L\alpha}) - {\cal O}_{ij}({\bf r} -{\bf r}_{S\alpha}) \right ] {\hat \nu_{\alpha j}} \;.
\end{equation}
The divergence at short distances is eliminated through the
definition: ${\cal O}_{ij}(|{\bf r}|  \leq a_{L,S}) = \delta_{ij}/\zeta_{L,S}$,
where  $\zeta_{L,S} = 6 \pi \eta a_{L,S}$.

The dynamics of an \emph {extended} body in Stokes flow
follows from translation of the hydrodynamic
centre and rotations about the hydrodynamic centre.
(The hydrodynamic centre refers to the point about which
the net hydrodynamic torque on the body vanishes; it
thus plays the same role in Stokes flow as the centre
of mass in inertial dynamics.) For the problem of a rigid dumbbell in an external flow, the 
hydrodynamic centre is obtained from
\begin{equation}
{\bf r }^C = \frac{\zeta_L {\bf r}_L + \zeta_S {\bf r}_S}{\zeta_L + \zeta_S} = \frac{a_L {\bf r}_L + a_S {\bf r}_S}{a_L + a_S}.
\end{equation}
The equations of motion for the translation and rotation of the hydrodynamic centre follow from
\begin{eqnarray}
\partial_t {\bf r}_\alpha^C &=&  v_0 {\hat \nu}_\alpha + \frac{1}{\bar \zeta} \sum_{\beta \neq \alpha} {\bf F}_{\alpha \beta} + \Gamma_\alpha(t) \nonumber \\
\omega_\alpha &=& \frac{1}{\zeta_R}  \sum_{\beta \neq \alpha} \tau_{\alpha \beta} + \Gamma^R_\alpha(t) \;,
\end{eqnarray}
where the angular velocity describing rotations about the hydrodynamic centre is defined by
\begin{equation}
\partial_t {\hat \nu}_\alpha ={\hat \nu}_\alpha \times \omega_\alpha
\end{equation}
and
\begin{equation}
{\bar \zeta} = (\zeta_L +\zeta_S)/2, \zeta_R = \ell^2 {\bar \zeta} \;.
\end{equation}
The random forces $\Gamma_\alpha$ and $\Gamma^R_\alpha$ lead to 
diffusion at large length scales~\cite{aparnaPNAS09}. 

The forces $F_{\alpha\beta}$ and the torques $\tau_{\alpha\beta}$ 
arise from hydrodynamic couplings between swimmers.
An isolated swimmer is propelled at speed
\begin{equation}
v_0 = -\frac{f \Delta a}{8\pi\eta \ell {\bar a}}
\end{equation}
with velocity $v_0 {\hat \nu}$. This velocity arises purely as a consequence 
of the fact that the hydrodynamic and the geometric centers do not coincide.
For symmetric swimmers, this velocity is zero and the swimmer is a ``shaker''
as opposed to a ``mover''.

The interactions between swimmers can be calculated in the dilute limit
by a multipole expansion, yielding
\begin{eqnarray}
F_{12} \approx 2 f {\bar a} \ell \left [3( \hat r_{12} \cdot {\hat \nu}_2)^2 -1 \right ]\frac{\hat {\bf r}_{12}}{r^2_{12}}
\end{eqnarray}
for the hydrodynamic force exerted by the $\beta$th swimmer on the
$\alpha$th one. In addition, expressions for 
the hydrodynamic torque between swimmers in the dilute limit
can be derived and are presented in Ref.~\cite{aparnaPNAS09}.

The hydrodynamic force decays as $1/r_{12}^2$,
as follows from its dipole character. The torque
consists of two terms: One is nonzero even for shakers
and aligns swimmers regardless of their polarity. The
second term vanishes for symmetric swimmers, serving
to align swimmers of the same polarity. A more detailed
discussion of the structure of the forces and torques for
pushers and pullers is available in Ref.~\cite{aparnaPNAS09}.

\subsection{Multiple Swimmers}
The coarse-grained version of the many swimmer problem is
defined  through the following local fields\cite{aparnaPRL08, aparnaPRE08}. First, we must have
a density of active particles, defined microscopically in terms of
\begin{equation}
{c({\bf r},t)} = \langle \sum_\alpha \delta({\bf r} - {\bf r}_\alpha^C(t))\rangle\;.
\end{equation}

Second, in the case in which the axes of a fore--aft asymmetric swimmer
are largely aligned along a common direction -- as in a magnet -- we can define 
a local field describing polar order in the following way:
\begin{equation}
{\bf P}({\bf r},t) = \frac{1}{c({\bf r},t)} \langle \sum_\alpha {\hat \nu}_\alpha \delta({\bf r} - {\bf r}_\alpha^C(t))\rangle\;.
\end{equation}

Third, and finally, when one considers ensembles of interacting swimmers,
we must also consider the possibility of additional and more subtle
\emph{macroscopic} variables representing orientational order.
An ensemble of individual particles, each aligned,
on average, along a common \emph{axis}, is familiar in soft
condensed matter physics. Such systems are referred
to as nematics and the ordering as nematic ordering.
In such \emph {nematic} order, the alignment is along a common axis but a
vectorial direction is not picked out.  
 
Orientational order in the nematic phase is generally described by a second-rank, 
symmetric traceless tensor $Q_{\alpha\beta}({\bf x}, t)$, 
defined in terms of  the  second 
moment of the microscopic orientational distribution 
function. This (order-parameter) tensor can  be expanded as 
\begin{equation}
Q_{\alpha \beta}=\frac{3}{2}S\left(n_{\alpha}n_{\beta} - \frac{1}{3}\delta_{\alpha \beta}\right) + 
\frac{1}{2}T\left(l_{\alpha}l_{\beta} - m_{\alpha}m_{\beta}\right) \;.
\label{Qtensor}
\end{equation}

The three principal axes of this tensor, obtained by diagonalizing $Q_{\alpha\beta}$ 
in a local frame, specify the direction of nematic ordering ${\bf n}$, the 
codirector ${\bf l}$ and the joint normal to these, labeled by ${\bf m}$. The principal values $S$ and $T$ represent the strength of ordering in the direction of ${\bf n}$ and ${\bf m}$, quantifying, respectively, the degree of uniaxial 
and biaxial nematic orders. 

In thermal equilibrium, the energetics of $Q_{\alpha\beta}$ is calculated from
a Ginzburg-Landau functional, first proposed by de Gennes, based on an 
expansion in rotationally invariant combinations of $Q_{\alpha\beta}$ 
and its gradients~\cite{degennes}. The Ginzburg-Landau-de Gennes functional $F$ is
\begin{eqnarray}
F &=& \int d^3{\bf x} [\frac{1}{2}ATr{\bf Q}^{2} + \frac{1}{3}BTr{\bf Q}^{3} + 
\frac{1}{4}C(Tr{\bf Q}^{2})^{2} \\\nonumber
&+& E^{\prime}(Tr{\bf Q}^{3})^{2} + \frac{1}{2}L_{1}(\partial_{\alpha}Q_{\beta\gamma})(\partial_{\alpha}Q_{\beta\gamma})]. 
\label{frener}
\end{eqnarray}
 
Here, $A = A_{0}(1 - T/T^{*})$ $T^{*}$ denoting the supercooling transition 
temperature, $A_{0}$ is a constant, $L_{1}$  is an elastic 
constant and $\alpha, \beta, \gamma$ denote the 
Cartesian directions. Other elastic terms can also be included; this simple
approximation corresponds to what is called the one Frank constant 
approximation.

Two simplifications are possible and often convenient. First, we may assume
uniaxial rather than biaxial order, since this is by far the more common
form of ordering.  In the ordered nematic state, the average orientation
occurs along a direction $\hat{n}$. This is the nematic
director, defined to be a unit vector.
We can  thus  work within a description in which
the components of $Q_{\alpha\beta}$ are written out in terms of the 
components of the nematic director $\bf n$. 
This gives us
\begin{equation}
Q_{\alpha \beta} =\frac{3S}{2} \left (n_i n_j - \frac{1}{3} \delta_{ij} \right ) \;.
\end{equation}
The energetics of small deviations from the aligned state is obtained,
in this representation, from a Frank free energy
appropriate to uniaxial nematics:
\begin{equation}
f_{FO} = \frac{1}{2} K_1 (div~{\bf n})^2 + \frac{1}{2} K_2 ({\bf n}~\cdot curl~
{\bf n})^2 + \frac{1}{2} K_3 ({\bf n} \times curl~{\bf n})^2 \;.
\end{equation}
Here $K_1$ is the splay elastic modulus, associated with a  splay deformation $\nabla \cdot {\bf n}$,
K$_2$ is the twist elastic modulus and  $K_3$ is the bend elastic modulus.
The Frank constants $K_1,K_2$ and $K_3$ have the dimensions of a force and
can be represented as the
ratio of an energy to a length scale. We can assume 
$K_i \sim k_B T_c/a$ where $a$ is a molecular length of order
$1\,nm$. Note that this description uses three elastic constants (which
can be reduced to the single elastic coefficient of Eq.~\ref{frener}
by assuming that $K_1=K_2=K_3$.

With this background, the definition of a local field
representing nematic order follows from~\cite{degennes},
\begin{equation}
Q_{ij} = \frac{1}{c({\bf r},t)} \langle \sum_\alpha ({\hat \nu}_{\alpha i}{\hat \nu}_{\alpha j}  - \frac{1}{3}\delta_{ij}) \delta({\bf r} - {\bf r}_\alpha^C(t))\rangle \;.
\end{equation}

\section{Hydrodynamic Approaches to Active Matter}
\label{sec:activelc}

A class of questions relating to the modeling of
active matter is concerned with (a) whether forms of collective
ordering are at all possible in ensembles of 
interacting active particles and (b) whether
such ordering, if assumed to preexist, can be shown to
be stable against fluctuations.  Our definition of the single swimmer associated a direction with the swimming motion, the direction of the axis formed by connecting, say, the center of the small sphere to the center of the large sphere. The question is thus whether the axes or orientations of different swimmers can be aligned as a consequence of their interaction.

In subsections below, the problem of deriving equations of motion
for the nematic order parameter field and the construction of a stress
tensor appropriate to a nematic fluid are briefly examined. Results for the
active nematic are summarized and the significance of these
results for the rheological properties of active matter briefly
outlined.

\subsection{Equations of Motion: Fluid and Nematic}

Identifying conservation laws and broken symmetries is
the crucial first step in constructing hydrodynamic equations of motion for the relevant fields in the problem.  For both conserved and hydrodynamic fields, 
the relaxation of long wavelength fluctuations proceeds slowly, with the relevant timescales for the relaxation of the fluctuation diverging as the 
wavelength of the perturbation approaches infinity.

For a simple fluid with no internal structure, the conservation 
laws for the local energy $\epsilon$, the density $\rho$ and
the three components of the momentum density ${\bf g}_i$ are
\begin{eqnarray}
\frac{\partial \epsilon}{\partial t} 
&=& -\nabla \cdot  j^\epsilon \nonumber \;,\\
\frac{\partial  \rho}{\partial t} &=& -\nabla \cdot g \nonumber \;,\\
\frac{\partial {\bf g}_i}{\partial t} &=& -\nabla_j \pi_{ij} \;,
\end{eqnarray}
where $j^\epsilon$ is the energy current and
$\pi_{ij}$ is the momentum current tensor, related to the
stress tensor. The conserved momentum density itself acts as a current 
for another conserved density, the mass (equivalently, number)
density, a relation which is responsible for sound waves in fluids.

The hydrodynamic description of fluids with internal order (such
as the nematic or polar fluid) must account for additional hydrodynamic
modes arising out of the fact that the ordering represents a broken
symmetry.  For small deviations from equilibrium, one derives an equation for
entropy generation and casts it in terms of
the product of a flux and a force. Such fluxes must vanish at thermodynamic
equilibrium. Close to equilibrium,
it is reasonable to expect that fluxes should have a smooth expansion in terms
of forces. 

As an illustrative example, consider  the simple fluid in the absence of dissipation.  We have,
with $\bf u$ the velocity field,
\begin{eqnarray}
{\bf g} &=& \rho {\bf u} \nonumber \;,\\
\pi_{ij} &=& p \delta_{ij} + u_j g_i = -\sigma_{ij} + \rho u_i u_j \nonumber \;,\\
{\bf j}^\epsilon &=& (\epsilon + p){\bf u} = \left ( \epsilon_0 + p + \frac{\rho u^2}{2} \right )\;.
\end{eqnarray}
The mass conservation equation is just
\begin{equation}
\frac{\partial \rho}{\partial t} = -\nabla \cdot (\rho {\bf u}),
\end{equation}
while the momentum conservation equation is
\begin{equation}
\frac{\partial {\bf g}_i}{\partial t} =  
\frac{\partial \rho {\bf u}_i}{\partial t} =  
-\nabla_j \pi_{ij} 
= -\nabla_i p - \nabla_j (\rho u_i u_j) \;.
\end{equation}
This is Euler's equation, usually written as
\begin{equation}
\frac{\partial {\bf u}}{\partial t} + ({\bf u} \cdot \nabla) {\bf u}
= \frac{-1}{\rho} \nabla p \;.
\end{equation}
The dissipative contribution to the stress tensor 
is accounted for by adding a term $\sigma^{'}_{ij}$ to
the stress tensor,
\begin{equation}
\pi_{ij} = p \delta_{ij} + \rho u_i u_j - \sigma^{'}_{ij} \;.
\end{equation}
Dissipation can only arise from velocity gradients, since any
constant term added to the velocity can be removed via a
Galilean transformation.
The dissipative coefficient coupling the stress tensor to the velocity gradient
is most generally a fourth rank tensor,
\begin{equation}
\sigma^{'}_{ij} = \eta_{ijkl} \nabla_k u_l \;.
\end{equation}
However, symmetry requires that
\begin{equation}
\sigma^{'}_{ij} = \eta(\nabla_i u_j + \nabla_j u_i - \frac{2}{3} \delta_{ij}
\nabla \cdot {\bf u} ) + \zeta \delta_{ij} \nabla \cdot {\bf u} \;.
\end{equation}
This gives the Navier-Stokes equations. Assuming  incompressibility, we have
\begin{equation}
\frac{\partial \rho}{\partial t} = 0 
= -\nabla \cdot (\rho {\bf u}) = - \nabla \cdot {\bf u} \;.
\end{equation}
Thus,
\begin{equation}
\rho \frac{\partial {\bf u}}{\partial t} + \rho ({\bf u} \cdot \nabla) {\bf u}
= -\nabla p + \eta \nabla^2 {\bf u} \;,
\end{equation}
along with the constraint $\nabla \cdot {\bf u} = 0$. The velocity field thus
has purely transverse components.

How is this to be generalized for a nematic fluid? 
First, for a nematic fluid, we must have an equation of motion for the
director $\bf n$ (or, equivalently, for the ${\bf Q}_{\alpha\beta}$ tensor)
in addition to the equations for the conservation of matter, momentum and
energy~\cite{degennes,stark}.  Second, distortions of configurations of the nematic order 
parameter field also contribute to the stress tensor of the system~\cite{degennes,stark}.

The director is aligned with the local molecular field in equilibrium;
local distortions away from the molecular field direction 
must relax in order to minimize the free energy. The local molecular
field is defined in the equal Frank constant approximation as
\begin{equation}
h_i = K \nabla^2 n_i \;.
\end{equation}
Also,  the director does not change under rigid translations at constant 
velocity. Thus,
the leading coupling of $\bf n$ to $\bf u$ must involve gradients of $\bf n$.
This can then be written as
\begin{equation}
\frac{\partial n_i}{\partial t} - \lambda_{ijk} \nabla_j u_k +X^{'}_i = 0 \;,
\end{equation}
where $X^{'}$ is the dissipative part of the current.
This dissipative part can be written as 
\begin{equation}
X^{'}_i = \delta^T_{ij} \frac{1}{\gamma} h_j \;,
\end{equation}
where $\gamma$ is a dissipative coefficient and the projector isolates
components of the fluctuation in the plane perpendicular to the 
molecular field direction.

The constraint ${\bf n} \cdot \partial {\bf n}/\partial t = 0$ implies that there are
only two independent components of the tensor $\lambda_{ijk}$. These
can be taken to be symmetric and antisymmetric, defining
\begin{equation}
\lambda_{ijk} = \frac{1}{2} \lambda(\delta^T_{ij} n_k + \delta^T_{ik} n_j)
+  \frac{1}{2} \lambda_2(\delta^T_{ij} n_k - \delta^T_{ik} n_j) \;,
\end{equation}
where
\begin{equation}
\delta^T_{ij} = \delta_{ij} - n_i n_j \;.
\end{equation}
Under a rigid rotation,
\begin{equation}
\frac{\partial n_i}{\partial t} = {\bf \omega} \times {\bf n}
= \frac{1}{2} ({\boldsymbol \nabla} \times {\bf u}) \times {\bf n} \;,
\end{equation}
mandating that the coefficient $\lambda_2$ of the antisymmetric part must be -1.

Thus the final equation of motion for the director, the Oseen equation,
takes the form~\cite{degennes,stark}
\begin{equation}
\partial_t n_i + {\bf v} \cdot \nabla n_i + \omega_{ij} n_j = 
\delta^{T}_{ij} \left (\lambda u_{ij} n_j + \frac{h_i}{\gamma} \right ) \;.
\end{equation}
where $u_{ij} = (1/2)(\partial_i u_j + \partial_j u_i)$.

The stress tensor of the nematic consists of three parts. The first
is the thermodynamic pressure $p$, while the second is the 
viscous stress, given by the tensor
\begin{eqnarray}
\sigma_{ij}^{'} &=& \alpha_1 n_i n_j n_k n_l A_{kl} + \alpha_2 n_j N_i
+ \alpha_3 n_i N_j + \alpha_4 A_{ij} 
+ \alpha_5 n_j n_k A_{ik} 
+ \alpha_6 n_i n_k A_{jk} \;,
\end{eqnarray}
constructed from symmetry allowed terms,
where ${\bf A}$ represents the symmetric part of the velocity gradient as before
and the vector $N_i = {\dot n}_i + \frac{1}{2}({\hat {\bf n}} \times curl {\bf u})_i$
is the change of director with respect to the background fluid. One relation,
due to Parodi,
connects the coefficients $a_1 \ldots a_6$; there are thus 5 independent 
coefficients of viscosity in the nematic~\cite{degennes}. 

The third contribution to the stress tensor 
is the static (elastic) contribution
arising from deformations in the director field~\cite{degennes}, i.e.,
\begin{equation}
\sigma^{e}_{ij} = - \frac{\partial F}{\partial \nabla_j n_k} \nabla_i n_k \;.
\end{equation}
A more detailed discussion of the equation of motion of the nematic
order parameter and the stress tensor is available in Ref.~\cite{degennes}.

\subsection{Active Orientational Order and its Instabilities}
The central idea behind the modeling of the active
nematic is that out of thermal equilibrium, new  terms
enter the equation of motion for the nematic director
as well as  the stress tensor.\footnote{In this section, we follow the
treatments of Refs.~\cite{aditiPRL02, yhatPRL04} closely.}
These terms are more
``relevant''  than the terms mandated by thermodynamic
approaches, in the sense that their effects are
stronger at long wavelengths and at long times, i.e. in the thermodynamic limit. To demonstrate this, assume nematic or polar ordered particles,
given by a unit director field $\bf n$. The ``slow'' or hydrodynamic variables are (i) the concentration
fluctuations $\delta c({\bf r}, t)$, (ii) total (solute
and solvent) momentum density ${\bf g}({\bf r}, t) = \rho
{\bf u} ({\bf r},t)$ and broken symmetry variables whose
fluctuations $\delta n_\perp = {\bf n} - {\hat z}$.
For polar ordered suspensions, there is a non-zero
drift velocity $v_0 \bf n$.
The momentum density evolves via
\begin{equation}
\frac{\partial g_i}{\partial t} = -\nabla_j \sigma_{ij} \;,
\end{equation}
with $\sigma_{ij}$ the stress tensor. 

What is special about self-propelled particle systems is that the active
contribution to the stress tensor is proportional to the
nematic order parameter, i.e., 
\begin{equation}
\sigma^{a}_{ij} \propto \left (n_i n_j - \frac{1}{3} \delta_{ij} \right ) \;.
\end{equation}
Pascals law is thus violated in the active nematic, since it has  a non-zero
deviatoric stress. Also, crucially, terms obtained from the standard near-equilibrium
analysis are of higher order, since they involve gradients of the nematic order
parameter. 

This result
can be derived from a microscopic calculation using the
simple model for a single swimmer discussed above, by simply
Taylor expanding the dipole term which appears in the force density,
about the hydrodynamic centre. This yields
\begin{equation}
\sigma^{a}_{ij} = \frac{a_L + a_S}{2} f c({\bf r},t) \left (n_i n_j - \frac{1}{3} \delta_{ij} \right )\;,
\end{equation}
where  $a_L$ and a$_S$ were defined earlier in the
context of the single swimmer.
We assume an initially aligned state in which the director points, on average,
along the $\hat{z}$ direction. Fluctuations away from this are given by
$\delta {\bf n}_\perp = \hat{n} - \hat{z}$.
Fluctuations about the averaged value can be parameterized by
\begin{equation}
\delta{\bf n} = {
\left (\begin{array}{c}
\delta n_x\\
\delta n_y\\
1\\
 \end{array}\right ) \;.
}
 \end{equation}
Also, we can expand the velocity field $\bf u$ in terms of fluctuations about its
mean value $u_0$,
\begin{equation}
\delta{\bf u} = {
\left (\begin{array}{c}
\delta u_x\\
\delta u_y\\
u_0 + \delta u_z\\
 \end{array}\right ) \;.
}
 \end{equation}
 Finally, we must allow for concentration fluctuations, via
 \begin{equation}
 c({\bf r},t) = c_0 + \delta c({\bf r},t) \;.
 \end{equation}
Using the equation of motion for the velocity field, 
defined through ${\bf u} = {\bf g}/\rho$,
we get
\begin{equation}
\partial_t {\bf u} = -\nabla_j \left [\frac{(a_L + a_S)}{2} (\frac{f}{ \rho}) (c_0 + \delta c({\bf r},t)) (n_i n_j - \frac{1}{3} \delta_{ij}) \right ] \;.
\end{equation}
This then gives, once we insert the expressions for 
small fluctuations and retain terms to the lowest order
 \begin{eqnarray}
 \partial_t \delta u_x &=& \frac{(a_L + a_S)}{2}  (\frac{f}{ \rho}) c_0 \partial_z \delta n_x \nonumber \\
 \partial_t \delta u_y &=&   \frac{(a_L + a_S)}{2}  (\frac{f}{ \rho}) c_0\partial_z \delta n_y \nonumber \\
 \partial_t \delta u_z &=&  \frac{(a_L + a_S)}{2}  (\frac{f}{ \rho}) c_0(\partial_x \delta n_x  +  \partial_y \delta n_y)  +  \frac{(a_L + a_S)}{2}  (\frac{f}{ \rho}) \partial_z \delta c({\bf r},t) \;,
 \end{eqnarray}
which we can write as
\begin{eqnarray}
 \partial_t \delta {\bf u}_\perp &\simeq& w_0\partial_z \delta {\bf n}_\perp \nonumber \\
 \partial_t \delta u_z &\simeq& w_0\nabla_\perp \cdot \delta {\bf n}_\perp  + \alpha\partial_z \delta c({\bf r},t) \;,
 \end{eqnarray}
where $\alpha \simeq f a/\rho$ and $w_0 \simeq c_0 a$ and
$a \simeq \frac{(a_L + a_S)}{2}$.
We must now impose incompressibility, which requires that
\begin{equation}
\nabla_\perp \delta {\bf u}_\perp + \partial_z u_z = 0 \;.
\end{equation}
In Fourier space, this is
\begin{equation}
i {\bf q}_\perp \cdot \delta {\bf u}_\perp({\bf q},t)  + i q_z  u_z(q_z,t) = 0 \;.
\end{equation}
Incompressibility implies that the velocity field must be purely transverse, a condition that is easily imposed by using the transverse projection operator
\begin{equation}
P^T_{\alpha \beta} = \left (\delta_{\alpha \beta} - \frac{q_\alpha q_\beta}{q^2} \right ) \;.
\end{equation}
Operating this projection operator on a vector field
isolates those parts of the field which have no longitudinal
component. Doing this yields
\begin{eqnarray}
\partial_t \delta u_x &=& -iw_0(1 - \frac{q_x^2}{q^2})  q_z \delta n_x - \frac{q_x q_y }{q^2} q_z \delta n_y 
- i\alpha\frac{q_x q_z }{q^2} (q_x \delta n_x + q_y \delta n_y + q_z \delta c) \nonumber \\
\partial_t \delta u_y &=& -iw_0 (- \frac{q_x q_y}{q^2})  q_z \delta n_x + ( 1 - \frac{q^2_y }{q^2})  q_z \delta n_y \nonumber\\
&-& i\alpha\frac{q_x q_z }{q^2} (q_x \delta n_x + q_y \delta n_y + q_z \delta c) \;,
\end{eqnarray}
which can be written in the compact form
\begin{equation}
\frac{\partial \delta {\bf u}_\perp}{\partial t} =
-i w_0 q_z \left ({\bf 1} - 2 {\bf q}_\perp {\bf q}_\perp/q^2 \right )
\delta {\bf n}_\perp 
- i \frac{q_z^2}{q^2} \alpha ({\bf q}_\perp \delta c) \;.
\end{equation}
Number conservation implies
\begin{equation}
\frac{\partial \delta c}{\partial t} = - \nabla \cdot {\bf j} \;,
\end{equation}
where ${\bf j} = c v_0 {\bf n} = c v_0 (\delta {\bf n}_\perp + \hat{z})$,
so
\begin{equation}
\left ( \frac{\partial }{\partial t} 
+ v_0\frac{\partial }{\partial z}  \right ) \delta c
+ c_0 v_0 \nabla_\perp \cdot \delta {\bf n} = 0 \;.
\end{equation}
The equation of motion for polar-ordered particles whose orientation is described by
${\bf n}$ contains a term representing
advection by a mean drift, a term describing the consequences of a non-equilibrium osmotic pressure
and other terms familiar from our brief description of nematodynamics close to equilibrium  in the previous
section\cite{aditiPRL02},
\begin{equation}
\frac{\partial \delta {\bf n}_\perp}{\partial t} 
= -\lambda_1 v_0 \frac{\partial \delta {\bf n}_\perp}{\partial z} 
- \sigma_1 \nabla_\perp \delta c
+ \frac{1}{2}\partial( \frac{ \delta {\bf u}_\perp}{\partial z} 
- \nabla_\perp u_z)
+ \frac{1}{2} \gamma_2(\frac{\partial \delta {\bf u}_\perp}{\partial z} 
+ \nabla_\perp u_z) \;.
\end{equation}

Taking the curl of this equation as well as the equation for
for ${\bf u}_\perp$ gives coupled equations for the dynamics of twist
$\nabla \times {\bf n}_\perp$ and/or vorticity $\nabla
\times {\bf u}_\perp$.  Related results follow from taking
the divergence of these two equations,  resulting
in coupled equations for $\nabla \cdot {\bf n}_\perp$,
$\nabla \cdot {\bf u}_\perp$ and $\nabla \delta c$,
which are complicated but have also been analysed~\cite{aditiPRL02}. 
These coupled equations have been shown to possess wave-like solutions.

The results of these calculations are summarized in the following:
\begin{itemize}
\item A linearized treatment, ignoring viscosity, for the polar or apolar cases
at lowest order in wave number, yields propagating modes with a 
characteristic instability in the case of  purely apolar active particles.
\item Retaining viscosity, in the steady Stokesian limit where accelerations are
ignored,  polar and nematic orders at small  wave numbers are generically destabilized by 
a coupling of splay (for contractile particles) or bend (for tensile particles) 
modes to the hydrodynamic velocity field.  In this limit,
for nematic order, this instability has been referred to as a generic instability. (A physical
picture for the origins of this instability is provided in Ref.~\cite{sriramNJP07}.)
\item Any mechanism for introducing screening into the hydrodynamic interaction
will suppress  the instability, such as introducing boundaries into the 
system~\cite{aparnaPNAS09,sriramNJP07}
or applying a shear~\cite{muhuri}.
\item Number fluctuations in ordered collections of self-propelled particles are 
anomalously large~\cite{tonertuPRL95,tonertuPRE98}. The variance
$\langle (\delta N)^2\rangle$, scaled by the mean $N$, diverges as $N^{2/3}$ in three
dimensions in the linearized treatment of Ref.~\cite{aditiPRL02}. Physically, this is a consequence of the fact that (a) orientational
fluctuations (director distortions) produce mass flow and (b) such
fluctuations are large because director fluctuations arise from a broken
symmetry mode. 
\end{itemize}

In a more general context, activity provides new currents both for matter and 
momentum, beyond those which would be predicted from a theory based on
small perturbations away from thermodynamic equilibrium. It is these new currents which
are the source of the novel results indicated above. 

\subsection{Rheological Predictions}
The following predictions for the rheological properties of active
nematics are obtained from \cite{yhatPRL04} and the
discussion here follows that reference closely; see also \cite{tanniePRL06,ahmediPRE06,andyPRE09}.
The active contribution to force densities within the fluid follows from
\begin{equation}
F^a = \nabla \cdot {\boldsymbol \sigma}^a \;.
\end{equation}
The activity contributes to
traceless symmetric (deviatoric) stress through
\begin{equation}
\sigma^a - \frac{1}{3} Tr {\boldsymbol \sigma}^a
\simeq W_1 {\bf Q} + W_2 {\bf Q}^2 \;,
\end{equation}
where $W_1$ and $W_2$ reflect the strength of the elementary dipoles. The
full stress tensor has contributions from the fluid (${\boldsymbol \sigma}^v$) as well as the 
order parameter field (${\boldsymbol \sigma}^{OP}$),
giving
\begin{equation}
{\boldsymbol \sigma}= {\boldsymbol \sigma}^a + {\boldsymbol \sigma}^v + {\boldsymbol \sigma}^{OP} \;.
\end{equation}
The time-dependence of  ${\boldsymbol \sigma}$ is assumed to be slaved to the time-dependence
of the order parameter field $\bf Q$. The equation of motion for $\bf Q$, upon
linearization, should take the form
\begin{equation}
\frac{\partial {\bf Q}}{\partial t} = -\frac{1}{\tau} {\bf Q}
+ D\nabla^2 {\bf Q} + \lambda_0 {\bf A} +\ldots \;,
\end{equation}
where $\tau$ is an activity correlation time, $D$ is a diffusivity, $\lambda_0$ 
is a kinetic coefficient and higher order terms have been dropped in favour of
the lowest-order ones.
Using these, we can calculate the stress response to a shear in the $xy$ plane.
In Fourier space, this is
\begin{eqnarray}
\sigma_{xy}(\omega) &=& -\left [\eta_0 + 
\frac{(a+W)\lambda_0}{-i\omega + \tau^{-1}} \right ] A_{xy} \nonumber \\
&=& - \frac{G^{'}(\omega) - i G^{''}(\omega)}{\omega} i A_{xy} \;,
\end{eqnarray}
defining the storage and loss moduli  $G^{'}(\omega)$ and $G^{''}(\omega)$.
The important results which follow from this analysis are the
active enhancement or reduction $\eta_{act} \propto W \tau$
(depending on the sign of the parameter $W$)
of the effective viscosity at zero shear rate. The theory also
predicts strong viscoelasticity as $\tau$ increases.
For passive systems $W=0$. For active systems, $W$ is non-zero
and the storage modulus then behaves as
\begin{equation}
G^{'}(\omega\tau \gg 1) \simeq W \;.
\end{equation}
All these effects are expected to be enhanced if the transition is
to a polar ordered phase, rather than to a nematic. 

\section{Active Gels: Summary}
\label{sec:activegel}
A parallel line of activity, centered on models for specific biological
phenomena such as the symmetry-breaking motility exhibited by
beads coated with polymerizing actin and the dynamical topological
defect structures obtained in mixtures of motors and microtubules, 
outlines a description of active matter in terms of
\emph {active gels}~\cite{giomiPRL08,julicher07,krusePRL04,kruseEPJE05,joannyHFSP}.  

The philosophy of these approaches is the following: 
Rather than begin from a  microscopic model for a swimmer or
individual moving particle and then generalize from the microscopics
to realize symmetry-allowed equations of motion for the fluid velocity
field and for the local concentration of swimmers, one can start with a 
coarse-grained continuum model for a
physical viscoelastic gel which is driven by internally generated, non-equilibrium sources of
energy\cite{joannyHFSP}. The equations of motion for the stresses in this gel as well as for local order-parameter-like quantities describing, for example, polar order at a coarse-grained scale are constructed using the basic symmetries of the problem. The non-equilibrium character of this problem follows
from the fact that such equations of motion do not derive from an underlying free 
energy\footnote{A summary of these results can be found in \cite{julicher07},
from where most of this material is drawn.}.

The passive gel has a viscoelasticity whose simplest representation is via
a Maxwell model, exhibiting solid-like behaviour at short times and fluid-like
behaviour at long times. In this model, the deviatoric stress $\sigma_{\alpha \beta}$ 
is related to the strain rate tensor $u_{\alpha \beta} = \frac{1}{2} \left ( \partial_\alpha u_\beta
+ \partial_\beta u_\alpha \right )$, where $\bf u$ is the velocity field in the
gel, via
\begin{equation}
\frac{\partial \sigma_{\alpha \beta}}{\partial t} + \frac{\sigma_{\alpha \beta}}{\tau} = 2 E u_{\alpha \beta} \;,
\end{equation}
where $E$ is a shear modulus obtained at short times. The simple time derivative must be augmented by convective terms, as well as terms
representing the effects of local rotation of the fluid, to enforce Galilean invariance
and the appropriate \emph{frame independence}.

Polar order in such  gels is assumed to be weighted 
 by a free-energy-like expression obtained from the
theory of polar nematic liquid crystals. This takes the form:
\begin{equation}
{\cal F} = \int d{\bf r} \left [\frac{K_1}{2} (\nabla \cdot {\bf p})^2 +
\frac{K_2}{2} ({\bf p} \cdot (\nabla \times {\bf p}))^2 +
\frac{K_3}{2} ({\bf p} \times (\nabla \times {\bf p}))^2
+ k \nabla \cdot {\bf p} - \frac{h_\parallel^0}{2} {\bf p}^2
 \right ]\;.
\end{equation}

There are three Frank constants for splay, twist and bend, as in the nematic case. The (non-zero)
constant $k$ is permitted by the vectorial symmetry of the polar case. The amplitude of
local order is parametrized by the constant $h_\parallel^0$.

The hydrodynamic theory of active gels begins by identifying, along
classical lines, fluxes and forces. The hydrodynamic description contains
phenomenological parameters, called Onsager coefficients.
These fluxes are the mechanical stress
$\sigma_{\alpha\beta}$ associated with the mechanical behaviour of the
cell, the rate of change of polar order (the polarization) ${\dot {\bf P}}$ and
the rate of consumption of ATP per unit volume r. The generalized force
conjugate to the ATP consumption rate  is the chemical potential difference
$\Delta \mu$ between ATP and the products of ATP hydrolysis, while the
force conjugate to changes in the polarization is the local field ${\bf h}$, obtained
from the functional derivative of the free energy, i.e.~${\bf h} = -\delta {\cal F}/\delta {\bf p}$. The force
conjugate to the stress tensor is, as usual, the velocity gradient tensor
$\partial_\alpha u_\beta$. This can, as is conventionally done, be expanded into
its traceless symmetric, pure trace and antisymmetric parts. A similar
expansion can be made for the stress tensor.

The next step is to construct equations of motion for the deviatoric stress,
using the convected Maxwell model with a single viscoelastic relaxation time.
The equation must couple the mechanical stress and the polarization  field 
as well as include a term coupling activity to the stress. It takes
the form
\begin{equation}
2\eta u_{\alpha \beta} = \left (1 + \tau\frac{D}{Dt} \right ) \left \{ 
\sigma_{\alpha \beta} + \zeta \Delta \mu q_{\alpha \beta} 
+ \tau A_{\alpha \beta} - \frac{\nu_1}{2} (p_\alpha h_\beta
+ p_\beta h_\alpha - \frac{2}{3} h_\gamma p_\gamma \delta _{\alpha \beta})
\right \} \;,
\end{equation}
where the co-rotational derivative is
\begin{equation}
\frac{D}{Dt} \sigma_{\alpha \beta}  = \left ( \frac{\partial}{\partial t} + u_\gamma \frac{\partial}{\partial r_\gamma}  \right )\sigma_{\alpha \beta}
+ \left [\omega_{\alpha\gamma} \sigma_{\gamma \beta} +  \omega_{\beta \gamma} \sigma_{\gamma \alpha} \right ] \;;
\end{equation}
 the tensor $A_{\alpha \beta}$ describes geometrical nonlinearities arising out of generalizations of
the Maxwell model to viscoelastic fluids, and
$q_{\alpha\beta} = \frac{1}{2} ( p_\alpha p_\beta - \frac{1}{3} p^2 \delta_{\alpha \beta})$. The antisymmetric part of the
 stress tensor leads to torques on the fluid and is obtainable from
\begin{equation}
\sigma^a_{\alpha\beta} = \frac{1}{2} \left (p_\alpha h_\beta - p_\beta h_\alpha \right ) \;.
\end{equation}
The viscoelastic relaxation time is $\tau$, the coefficient $\nu_1$ describes the coupling between mechanical stresses and the polarization field, while the parameter $\zeta$ is the coefficient of active stress generation, acting to couple activity to the stress.
The second flux, defined from the rate of change of polarization is given
by
\begin{equation}
{\dot {\bf P}} = \frac{D {\bf P}}{Dt} \;.
\end{equation}
The Onsager relation for the polarization is
\begin{equation}
\frac{D}{Dt} p_\alpha  = \frac{1}{\gamma_1} h_\alpha + \lambda_1 p_\alpha \Delta \mu - \nu_1 u_{\alpha \beta} p_\beta
- {\bar \nu}_1 u_{\beta \beta} p_\alpha \;.
\end{equation}
These include several phenomenological parameters, such as 
the rotational viscosity $\gamma_1$ which characterizes dissipation from the rotation of the polarization
as well as the constants $\nu_1$ and ${\bar \nu}_1$.

Then, we must have an equation for the rate of consumption of ATP. This takes the form
\begin{equation}
r = \Lambda \Delta \mu + \zeta p_\alpha p_\beta u_{\alpha \beta} + {\bar \zeta} u_{\alpha \alpha} + \lambda_1 p_\alpha h_\alpha \;.
\end{equation}

These are simple but generic equations representing the basic symmetries of the problem\cite{joannyHFSP}. They can be shown to have
interesting and surprising consequences: an active polar gel can exhibit spontaneous motion as a consequence of a gradient in the
polar order parameter, as well as defects in the
polar ordering which are dynamic in character~\cite{julicher07}. 
These ideas  have been applied to the study of the motion of the cell lamellipodium and to the organization of microtubules
by molecular motors\cite{joannyHFSP}. 

The generality of these equations follows from the fact that they are motivated principally by symmetry considerations.
Thus, even though they describe intrinsically non-equilibrium  and highly nonlinear phenomena, for which the rules of constructing
effective, coarse-grained equations of motion for the basic fields are not as well developed as the theory for the relaxation
of  small perturbations about thermal equilibrium, the ``unreasonable effectiveness of hydrodynamics'' may well hold in
their favour.

\section{Conclusions}
\label{sec:conclusions}
This chapter has provided a brief review of the field of what is currently
called active matter. The emphasis has been on attempting to clarify  the basic ideas 
which have motivated the development of this field, rather than the details of the often
intricate and complex calculations implementing these ideas. Much recent and
important work, including numerical calculations -- illustrative references are Refs.~\cite{saintillan,marenduzzo,sanoop} --
has been omitted entirely for the sake of compactness. 

As indicated  in the introduction, the importance of this field  would appear to be that it might  suggest
ways of thinking about the response and dynamics of living systems, while providing a 
largely self-consistent framework for calculations.  Several non-trivial insights have already
been obtained from these calculations, particularly in the identification of the generic instability of
polar or nematically ordered states in the presence of the long-ranged hydrodynamic interaction,
the connection between microscopic models and their hydrodynamic limits, as well as a comprehensive
theory of active gels, generalizing ideas from nematic physics. The precise relationship between macroscopic,
symmetry-based hydrodynamic equations representing active nematics and  an underlying microscopic
theory has been substantially clarified, as in the work of Ref.~\cite{aparnaPNAS09} and references
cited therein. To what extent further developments 
in this field may aid the increasingly \emph {active} dialogue between physics and the engineering sciences 
on the one hand and the biological sciences on the other remains to be seen.

\vskip 1cm
{\em I thank Sriram Ramaswamy and Madan Rao for many enlightening and valuable
discussions concerning the physics of active matter. Conversations at various points of time with Cristina Marchetti, Tanniemola Liverpool, Jacques Prost, Karsten Kruse, Frank Julicher, David Lacoste, Ronojoy Adhikari,
P. B. Sunil Kumar, Aparna Baskaran and Jean-Francois Joanny have also helped to shape the material in this chapter. This work was supported by DST (India) 
through a Swarnajayanti Fellowship, by the DST Nanomission [Grant SR/S5/NM-
10/2006], by the Indo-French Centre for the Promotion of Advanced Research  (CEFIPRA) [Grant No. 3502] as well as by the PRISM project, IMSc. The hospitality of ESPCI and the Institut Henri Poncare is gratefully acknowledged.}

\end{document}